\documentclass[11pt]{article}
\usepackage{amsfonts}
\usepackage{amscd}
\usepackage{color}
\usepackage{graphicx}
\usepackage{latexsym}
\usepackage[english]{babel}

\newtheorem{rmq}{Remark}

\author{\bf Patrick Louodop, Hilaire Fotsin, Elie B. Megam Ngouonkadi\\
Laboratory of Electronics\\
Faculty of Science, Department of Physics,\\
University of Dschang, P.O. Box 67 Dschang, Cameroon\\
\\
\bf Samuel Bowong\\
Laboratory of Applied Mathematics, \\
Department of Mathematics and Computer Science, Faculty of Science, \\
University of Douala, P.O. Box 24157  Douala, Cameroon\\
\\
\bf Hilda A. Cerdeira$^\dag$\\
Instituto de F\'{i}sica Te\'{o}rica - UNESP, Universidade Estadual Paulista, \\
Rua Dr. Bento Teobaldo Ferraz 271, Bloco II, Barra Funda, 01140-070 S\~ao Paulo,
Brazil.\\
\\
$^\dag$ Corresponding author: Tel. +55 11 982880568\\
Email: cerdeira@ift.unesp.br}
\title{\textbf{Effective synchronization of a class of Chua's chaotic systems
using an exponential feedback coupling.}}
\date{}
\begin{document}
\maketitle
\baselineskip7mm
%\tableofcontents
\begin{abstract}
\noindent

In this work a robust exponential function based controller is designed to
synchronize effectively a given class of Chua's chaotic systems. The stability
of the drive-response systems framework is proved through the Lyapunov stability
theory. Computer simulations are given to illustrate and verify the method.
\end{abstract}

\section{Introduction}
\noindent

Shortly after Pecora and Carroll showed the possibility of synchronizing chaotic
elements [1], applications stretched out in many fields [2-4] giving rise to
interdisciplinary research, since this phenomenon appears in many systems in a
variety of ways [5-12]. Much research was done developing different strategies
in the quest of effective synchronization such as adaptive synchronization
[13-15], inverse synchronization [16], anti-synchronization [17-19] and so on
[20-24]. The robustness of many of these methods, as surprising as it may
appear, has already been demonstrated in many cases in the presence of noise,
perturbations or parameter mismatches [25-27].\\

For applications such as telecommunications, where the transmission of
messages is not possible unless transmitter and receiver are synchronized
[10-12], the investigation of new chaotic systems as well as the most effective
means of synchronization is always of great importance. Thus, mathematical
models [28,29], mechanical systems [30], electronic circuits [10, 13] are
continually built. One of the best known electronic circuits is the Chua's
oscillator [33,34]. Although Chua's circuit is one of the simplest circuits in
the literature, it has various complex chaotic dynamics properties which has
made it a topic of extensive study [31-34].  A modified version of the circuit
has also been topic of attention [31, 32, 35]. Its theoretical
analysis and numerical simulations agree very well with experimental results.
\\

Recently, some authors proposed a nonlinear controller in order to force
synchronization with the purpose of saving energy [22,23]. The nonlinear
controllers used are based on bounded nonlinear functions [22,23]. In this work
we apply the exponential function based nonlinear controller to achieve
synchronization between the drive-response systems when disturbances are
present. Our controller has certain properties which makes it more advantageous
to use it, such as: (1) it is easy to implement in practice; (2) it needs no
adaptation algorithm, hence its electrical circuit remains simple;(3) it is
faster than the synchronization based on fixed feedback gain which is usually
used.\\

This work is organized as follows: In section 2, the problem is formulated and
the assumptions are given. Section 3 presents the main results. We use Lyapunov
stability theory to study the robustness of our proposed controller. We show
that with this controller the drive and response systems
are practically synchronized - the
errors between the master system and the slave system do not tend to zero but
to a limit value. In this case, it is shown that the derivative
of Lyapunov function is contained in a closed domain to which the error between
master and slave system converge. Since
 the error is sufficiently small, using the principle of the "ultimate boundedness property", we arrive to the conclusion that the 
system is globally stable because the derivative of the Lyapunov function is negative definite. Ultimate boundedness is in particular compatible with local instability
about zero and implies global stability. This was demonstrated by
Z. Ding and G. Cheng in ref.[36]. They proposed a new criterion
of globally uniformly ultimate boundedness for discrete-time nonlinear systems which helps
to relax the condition of stability based on Lyapunov function. The same ideas were successfully applied by De La Sen and S. Alonso [37], while
in ref.[38], Bitsoris et al. work on the robust positive invariance and ultimate 
boundedness of nonlinear systems with unknown parameters and disturbances, where only their bounds of variance are known.
In section 4, numerical results are
presented and we compare the given scheme with that using the simple fixed gain
based controller. The conclusions are given in section 5.

\section{Formulation of the problem}
\noindent

In this paper, we study the master-slave synchronization of a class of Chua's
chaotic systems, represented in Figure 1 and described by the equations that follow. \\

The master system is given by:

\begin{equation}
\label{master} \left\{ \begin{array}{lcl}
\dot{x}_1\left(\tau\right)&=&\alpha
\left[x_2\left(\tau\right)-x_1\left(\tau\right)-R
f(x_1\left(\tau\right))\right]+ d\left(\tau\right),\\
\\
\dot{x}_2\left(\tau\right)&=&\beta\left[
x_1\left(\tau\right)-x_2\left(\tau\right)+Rx_3\left(\tau\right)\right],\\
\\
\dot{x}_3\left(\tau\right)&=&\gamma
\left[v\left(\tau\right)-x_2\left(\tau\right)\right],\end{array} \right.
\end{equation}
where $\tau$ is a dimensionless time, $x_i(t)$, $i=1,2,3$ are the state
variables, $v\left(\tau\right)$ is an external force and $\alpha, \beta, \gamma$
and $R$ are positive constant parameters of the system. The
function $f(x_1\left(\tau\right))$ represents the nonlinearity of the system and
$d\left(\tau\right)$ the disturbances. The function
$f(x_1\left(\tau\right))$ defines Chua's circuit, which is given by $f(x_1)=a_2
x\left(\tau\right)+0.5(a_1-a_2)(|x_1\left(\tau\right)+1|-|x_1\left(\tau\right)-1
|)$ [31-33], while the modified Chua's system is obtained using
$f(x_1\left(\tau\right))=a x_{1}^{3}$ [12] or
$f(x_1\left(\tau\right))=a_{1}(x_1-b)^{3}-a_{2}(x_1-b)+a_{3}$ [34]. The latter
represents the behavior of a tunnel diode [34]. For an autonomous system
$v\left(\tau\right)$ is constant and $d(\tau)=0$.\\

The slave system is given by:

\begin{equation}
\label{slave} \left\{ \begin{array}{lcl}
\dot{y}_1\left(\tau\right)&=&\alpha
\left[y_2\left(\tau\right)-y_1\left(\tau\right)-R
f(y_1\left(\tau\right))\right]+ U\left(\tau\right),\\
\\
\dot{y}_2\left(\tau\right)&=&\beta\left[
y_1\left(\tau\right)-y_2\left(\tau\right)+Ry_3\left(\tau\right)\right],\\
\\
\dot{y}_3\left(\tau\right)&=&\gamma
\left[v\left(\tau\right)-y_2\left(\tau\right)\right],\end{array} \right.
\end{equation}
where $y_i\left(\tau\right), i=1, 2, 3$ is the slave state variables and
$U\left(\tau\right)$ the feedback coupling.\\

Here we present a scheme to solve the synchronization problem for system (\ref{master}). That is to say,
if the uncertain system (\ref{master}) is regarded as the drive system, a
suitable response system should be constructed to synchronize it
with the help of  the  driving signal $x$.  In order to do so, we
assume that:
\begin{description}
  \item[(i)] There is a bounded region $\mathcal{U}\subset R^3$ containing
the whole basin of the drive system (\ref{master}) such that no orbit of
system (\ref{master}) ever leaves it.

  \item[(ii)] The disturbance $d(\tau)$ is  bounded by
an unknown positive constant  $D$, namely
\begin{equation} \label{perturb}
\|d\left(\tau\right)\| \leq D,
\end{equation}
where  $\|\cdot\|$ denotes the euclidian norm of a vector.
  \item[(iii)] All chaotic systems are supposed to be confined
to a limited domain, hence it exists a positive constant $L$
such that
  \begin{equation} \label{bound}
  \|f(y_1\left(\tau\right))-f(x_1\left(\tau\right))\| \leq
L\|y_1\left(\tau\right)-x_1\left(\tau\right)\|.
  \end{equation}

\end{description}

We shall now try to synchronize the systems described in (\ref{master}) and (\ref{slave}) designing
an appropriate control $U(\tau)$ in system (\ref{slave}) such that

 \begin{displaymath}
 \| y_i\left( \tau\right)- x_i \left(\tau\right)\| \leq r, \qquad \mbox{for
}\qquad \tau\rightarrow\infty,
\end{displaymath}
where $r$ is a sufficiently small positive constant.\\

Let us define the state errors between the transmitter and the receiver systems
as

 \begin{equation} \label{error}
  \begin{array}{lcl}
e_i \left( \tau\right) &=& y_{i}\left( \tau\right)- x_i\left( \tau\right),\qquad \mbox{with }\qquad i=1, 2, 3.
\end{array}
\end{equation}

and the feedback coupling as:

\begin{equation} \label{controller}
\begin{array}{lcl}
U(\tau)=-\varphi \left(\exp(k e_1\left( \tau\right))-1\right),
\end{array}
\end{equation}

where $\varphi$ and $k$ are positive fixed constants.\\

Introducing the definition of the systems (\ref{master}), (\ref{slave}) and
equation (\ref{controller}) into (\ref{slave}), the dynamics of the error states (\ref{error}) becomes:

\begin{equation}
\label{erdyn} \left\{ \begin{array}{lcl}
\dot{e}_1\left(\tau\right)&=&\alpha
\left[e_2\left(\tau\right)-e_1\left(\tau\right)-R\left(f(y_1)-
f(x_1)\right)\right]- d\left(\tau\right)-\varphi \left(\exp(k e_1)-1\right),\\
\\
\dot{e}_2\left(\tau\right)&=&\beta\left[
e_1\left(\tau\right)-e_2\left(\tau\right)+ Re_3\left(\tau\right)\right],\\
\\
\dot{e}_3\left(\tau\right)&=&-\gamma e_2\left(\tau\right),\end{array} \right.
\end{equation}

The problem is now reduced to demonstrating that with the
chosen control law $U(\tau)$, the error states $e_i, i=1, 2, 3$ in (\ref{erdyn}) are at
most a sufficiently small positive constant $r$, which will prove the
proposition.

\section{Main results}
\noindent

If we consider the master-slave chaotic systems (\ref{master}) and (\ref{slave}) with all the
aforementioned assumptions (\ref{perturb}) and with the exponential function based feedback
coupling given by the relation (\ref{controller}), we shall show that the overall system
will be  practically synchronized, i.e., $\| y_i\left( \tau\right)- x_i
\left(\tau\right)\| \leq r$, where $r$ is a sufficiently small positive constant
for large enough $\tau$.

In order to do so, let us consider  the following Lyapunov
function:
\begin{equation}\label{lya}
V =\displaystyle\frac{1}{2}\left[\displaystyle\frac{e_1^2}{\alpha}+\displaystyle\frac{
e_2^2}{\beta}+\displaystyle\frac{R~e_3^2}{\gamma}\right],
\end{equation}

Differentiating the function $V$ with respect to time yields

\begin{equation}\label{lyad1}
\begin{array}{lcl}
\dot{V} &=&2e_1e_2-e_1^2-e_2^2-R\left(f(y_1)- f(x_1)\right)
e_1-\displaystyle\frac{d\left(\tau\right)}{\alpha} e_1-\displaystyle\frac{\varphi}{\alpha} \left[\exp(k e_1)-1\right]e_1,\\
\\
&=&-\left(e_1-e_2\right)^2-R\left(f(y_1)- f(x_1)\right)
e_1-\displaystyle\frac{d\left(\tau\right)}{\alpha} e_1-\displaystyle\frac{\varphi}{\alpha} \left[\exp(k e_1)-1\right]e_1.
\end{array}\end{equation}

Expanding the exponential function as follows

\begin{equation}\label{expansion}
 \exp(k e_1)-1\simeq k e_1+\sum_{i=1}^{n}\displaystyle\frac{\left(k
e_1\right)^{2i}}{2i!}+\sum_{i=1}^{n}\displaystyle\frac{\left(k
e_1\right)^{2i+1}}{(2i+1)!}+\theta (e_1)+\zeta (e_1),
\end{equation}
where $\theta (e_1)$ and $\zeta (e_1)$ constitute the rest of the expansion in order greater than $n$ for odd part and 
for even part of the development respectively,
and substituting by the maximum value of the disturbance, $D$,
it follows that

\begin{equation}\label{lyad2}
\dot{V}\leq R L e_1^2+\displaystyle\frac{D}{\alpha} |e_1|-\displaystyle\frac{\varphi}{\alpha} \left(k
e_1+\sum_{i=1}^{n}\displaystyle\frac{\left(k
e_1\right)^{2i}}{2i!}+\theta (e_1)\right)e_1.
\end{equation}

Hence, we have

\begin{equation}\label{lyad3}
\dot{V} \leq \left(R L-\displaystyle\frac{\varphi}{\alpha} k \right)e_1^2+ \displaystyle\frac{1}{\alpha}\left(D+\varphi 
\sum_{i=1}^{n}\displaystyle\frac{\left(k
|e_1|\right)^{2i}}{2i!}+\left|\theta (e_1)\right|\right)|e_1|.
\end{equation}

\begin{equation}\label{lyad4}
\dot{V} \leq \left(R L-\displaystyle\frac{\varphi}{\alpha} k\right)e_1^2+ \displaystyle\frac{1}{\alpha}\left(D
+\varphi\sum_{i=1}^{n}\displaystyle\frac{\left(k r\right)^{2i}}{(2i)!}+\Theta (r)\right)r, \quad{where}\quad \Theta (r)\geq  \max{(\theta (e_1))}.
\end{equation}

Here we use $r$ as an upper bound for the error in each axis.
Then we see that the derivative of the Lyapunov function (\ref{lyad2}) is lower than that
in equation (\ref{lyad3}), which in turn is smaller than the one given by equation (\ref{lyad4}).
Thus expression (\ref{lyad4}) is maximized and  the radius of the close domain to which
the error is attracted is determined.
Defining

\begin{equation}\label{hyp}
\varphi\geq\displaystyle\frac{\alpha~R~L}{k} \qquad \mbox{and }\qquad   \phi=\left(\displaystyle\frac{D+\Theta (r)}{\alpha}
+\displaystyle\frac{R~L}{k} \sum_{i=1}^{n}\displaystyle\frac{\left(k r\right)^{2i}}{(2i)!}\right)r,
\end{equation}
one obtains

\begin{equation}\label{lyad5}
\begin{array}{lcl}
\dot{V} &\leq& - \psi e_1^2+ \phi \qquad \mbox{where }\qquad  \psi =\left|R L-\displaystyle\frac{\varphi}{\alpha} k\right|.
\end{array}
\end{equation}
Eq.~(\ref{lyad5}) is in principle a form of the ultimate boundedness property in the sense that if the error is
sufficiently small, then the system is globally stable because the upper-bound is negative [36].
From (\ref{lyad5}), it follows that if

\begin{center}
$\|e_1\|>\sqrt{\displaystyle\frac{\phi}{\psi}}$,\\
\end{center}
therefore, $\dot V\left(\tau\right)<0$, hence $V\left(\tau\right)$ decreases,
which implies that $\|e_1\|$ decreases. It then follows from standard invariance arguments as in Ref.[23]
that asymptotically for increasing time the error satisfies the following bound
\begin{center}
$\|e_1\|< C$,
\end{center}
for any $C > \sqrt{\displaystyle\frac{\phi}{\psi}}$.\\

So if $\phi$ is sufficiently small, the bound for the synchronization error will also be
sufficiently small. Therefore, the synchronization state error would be
contained within a neighborhood of the origin, as we wanted to prove. In addition as $V\left(\tau\right)$ decreases, then there exists a continuous and strictly increasing function $\varrho$ and a finite integer $\eta$, such that  $$V\left(e(\tau+\eta),\tau+\eta\right)-V\left(e(\tau),\tau\right)\leq -\varrho(\|e(\tau)\|)\quad{where}\quad e(\tau)=\left(e_1(\tau), e_2(\tau), e_3(\tau)\right).$$ Thus, the Lyapunov function respect the Theorem 3.1 in [36]
and then Eq.(\ref{erdyn}) is globally uniformly ultimate bounded near the origin.

\section{Numerical simulations}
\subsection{Chaotic systems}
\noindent

\begin{figure}[htp]
 \begin{minipage}[b]{16cm}
      \begin{center}
    \includegraphics[scale=0.35]{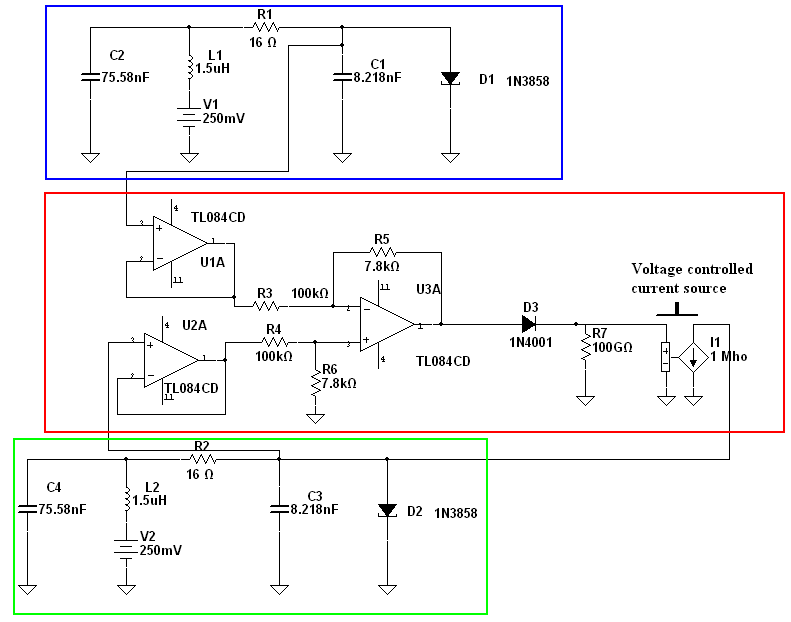}
    \caption{Scheme of the circuit for the whole system: transmitter (blue box), controller (red box) and receiver (green box).}
     \label{Fig.1}
     \end{center}
     \end{minipage}
     \end{figure}

\begin{figure}[htp]
 \begin{minipage}[b]{16cm}
      \begin{center}
    \includegraphics[scale=0.24]{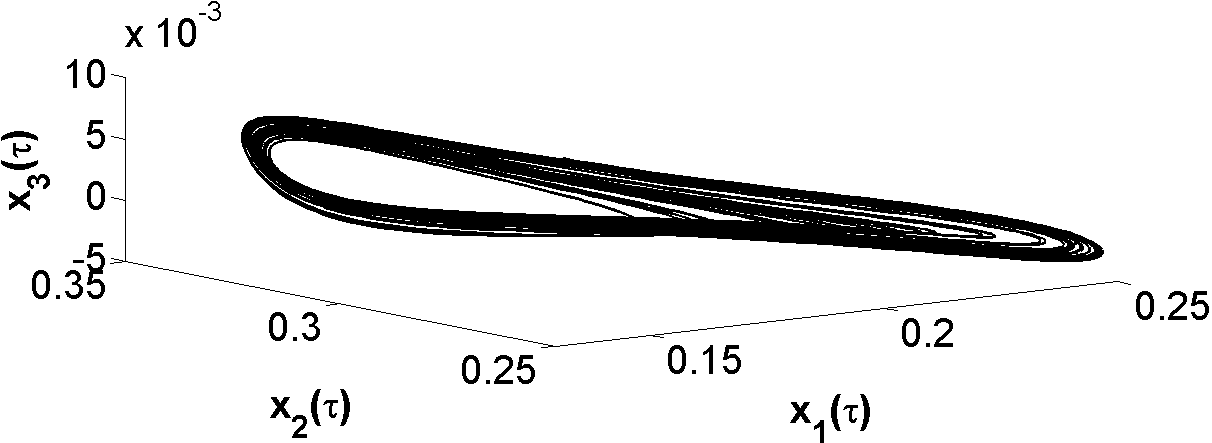}(a)
    \caption{3D-Chaotic attractor of the tunnel diode based modified Chua's
system.}
     \label{Fig.2}
     \end{center}
     \end{minipage}
     \end{figure}

In this section, we present some numerical results for the circuit shown in Figs~(\ref{Fig.1}), to illustrate the effectiveness of the
proposed scheme, where the three-dimensional -- tunnel diode based -- modified Chua's system
[35] is used as transmitter (blue box) and receiver (green box), the controller appears inside the red box.
With the initial conditions selected as
$(x_{1}(0),x_{2}(0),x_{3}(0))=(0.15, 0.27, 0.008)$ and
$(y_{1}(0),y_{2}(0),y_{3}(0))=(0.18, 0.24, 0.006)$ and with the given system's
parameters: $\alpha=2.507463$, $\beta=0.2985075$, $\gamma=0.20875$, $R=16$,
$e=0.250$, $a_1=1.3242872$, $a_2=0.06922314$, $a_3=0.00539$ and $b=0.167$ the
systems behave chaotically as shown in Fig.~(\ref{Fig.2}).
 The disturbances $d(\tau)$ are
given by the relation $d(\tau)=0.001
wgn(1,1,1)\left(x_1(\tau)+x_2(\tau)\right)$, $wgn(1,1,1)$ is Matlab white
gaussian noise generator.

\subsection{Simulation results and discussion}
\noindent

\begin{figure}[htp]
 \begin{minipage}[b]{16cm}
      \begin{center}
    \includegraphics[scale=0.22]{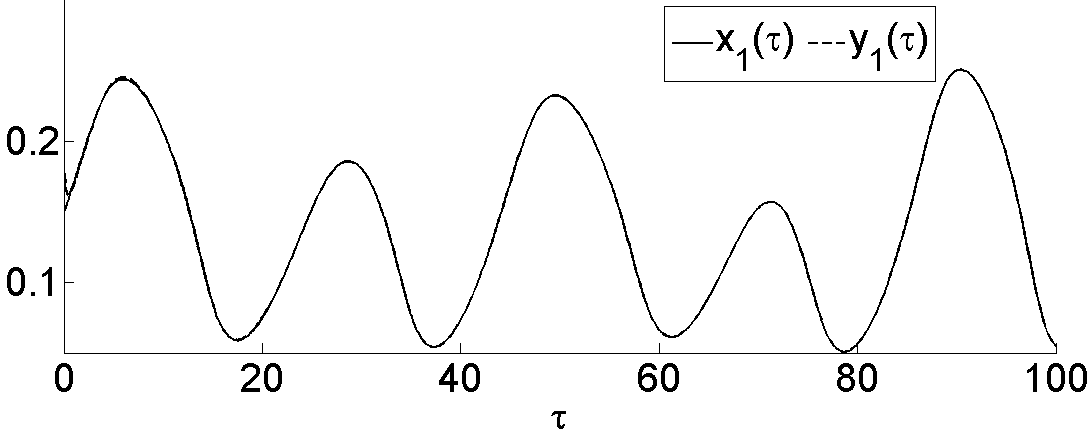}(a)
    \includegraphics[scale=0.22]{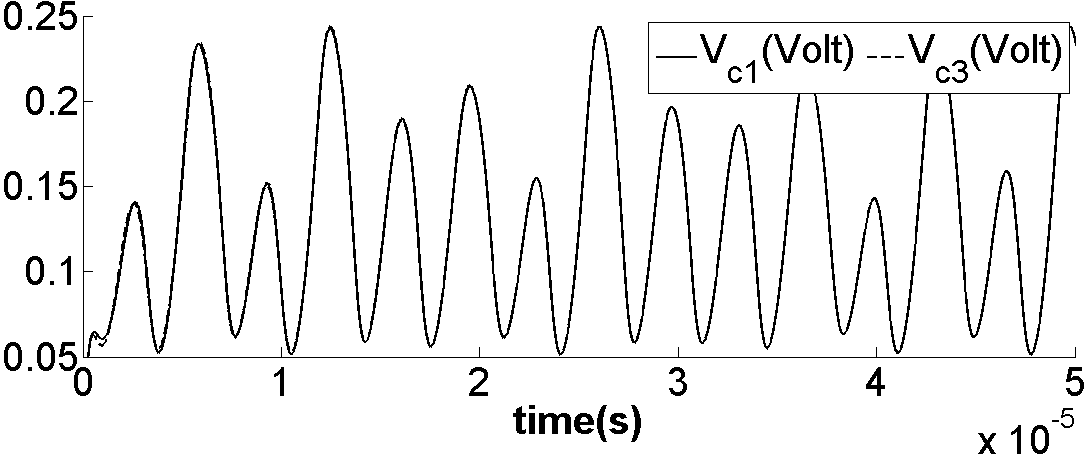}(b)
    \includegraphics[scale=0.22]{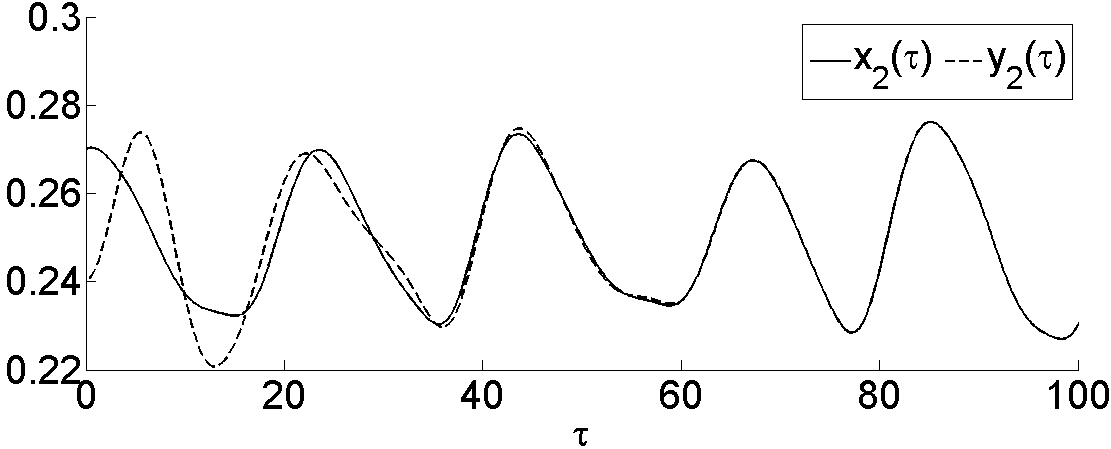}(c)
    \includegraphics[scale=0.22]{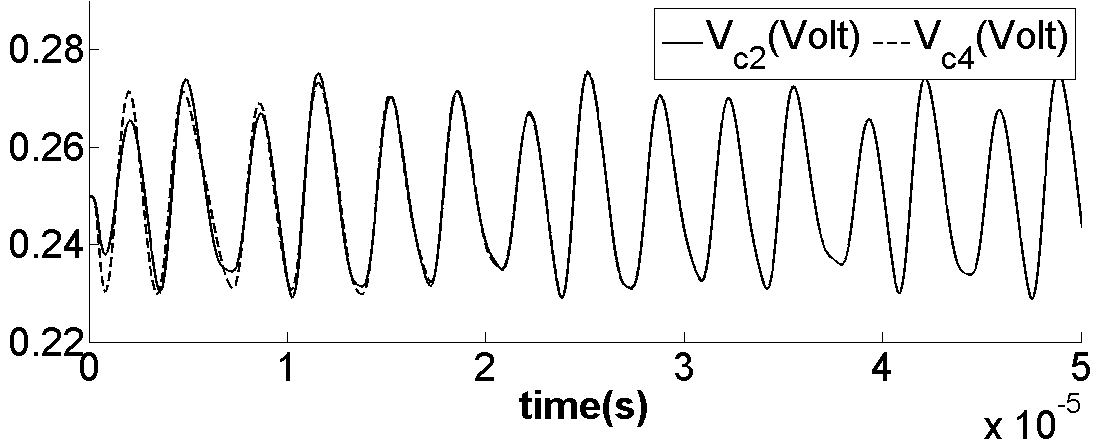}(d)
    \includegraphics[scale=0.22]{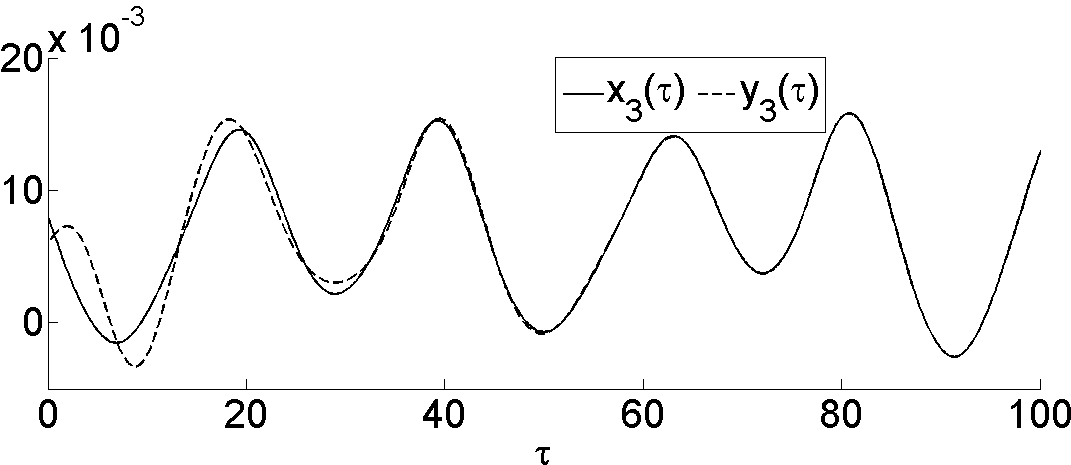}(e)
    \includegraphics[scale=0.22]{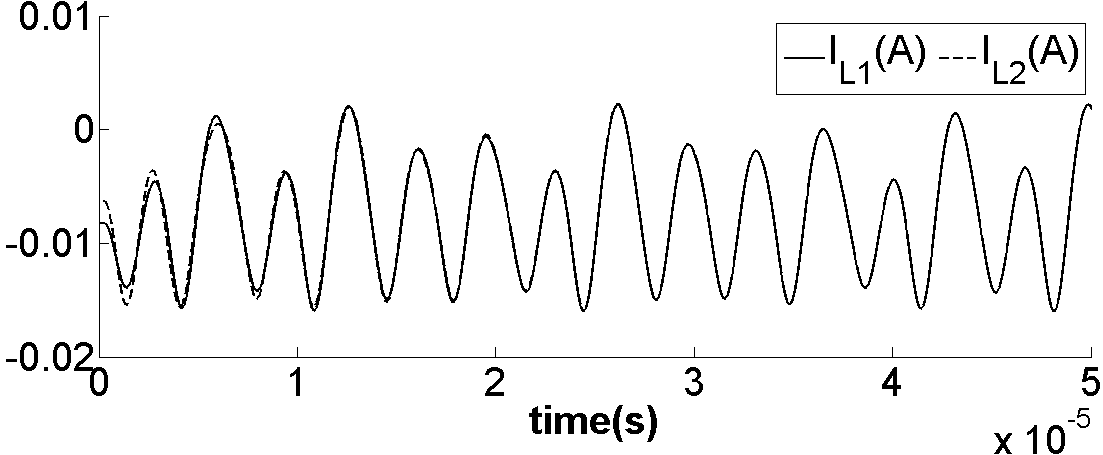}(f)
    \caption{Time evolution of the master system (solid lines) and slave system
(dashed lines) from Matlab simulations (Left) and Pspice simulations (Right).}
     \label{Fig.3}
     \end{center}
     \end{minipage}
     \end{figure}

\begin{figure}[htp]
 \begin{minipage}[b]{16cm}
      \begin{center}
    \includegraphics[scale=0.22]{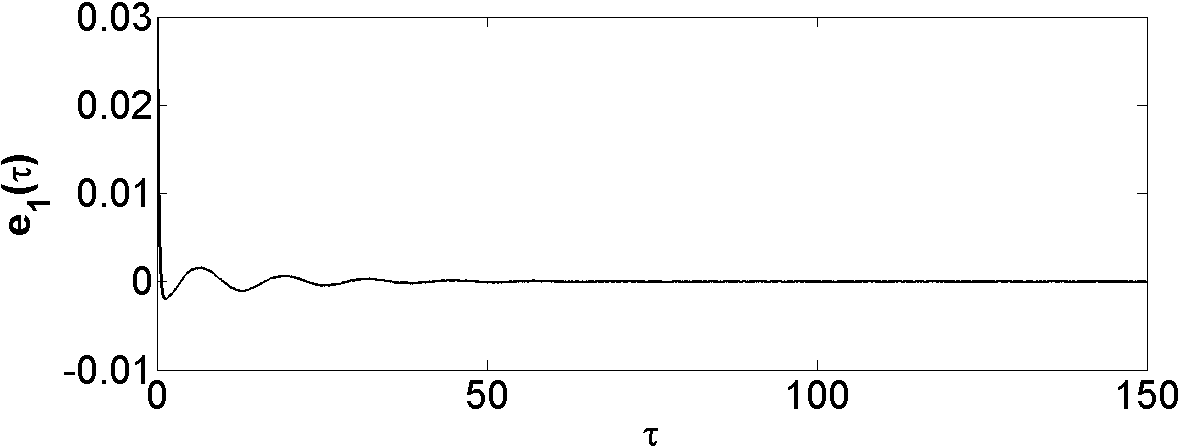}(a)
    \includegraphics[scale=0.22]{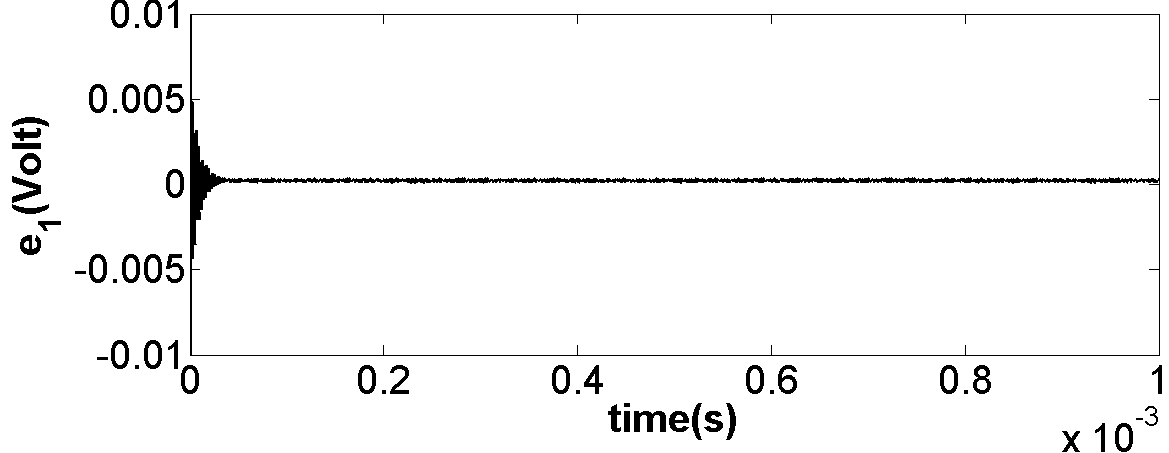}(b)
    \includegraphics[scale=0.22]{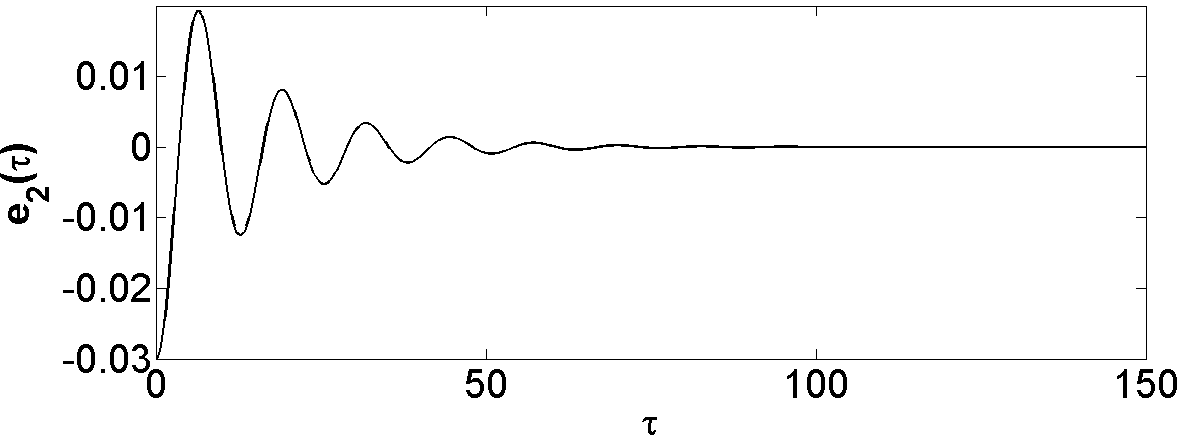}(c)
    \includegraphics[scale=0.22]{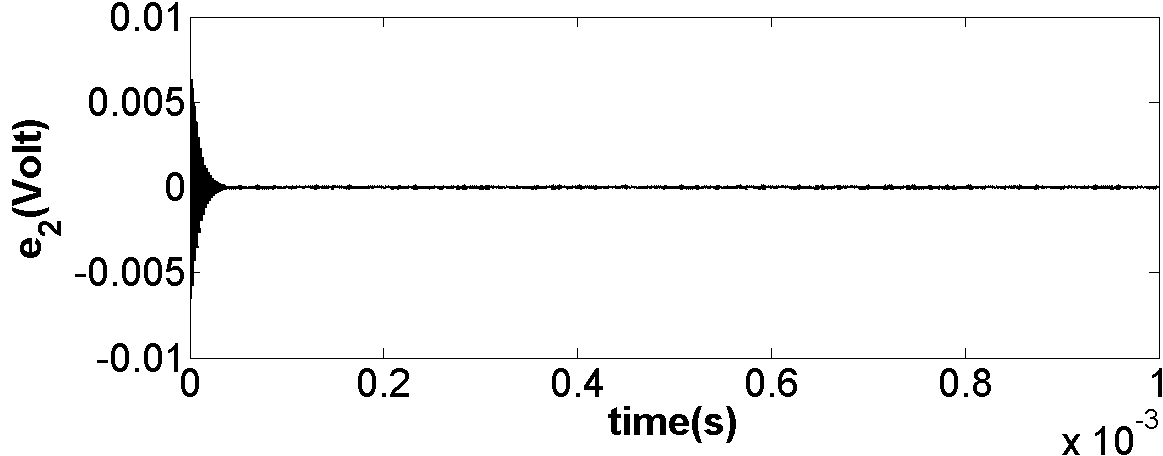}(d)
    \includegraphics[scale=0.22]{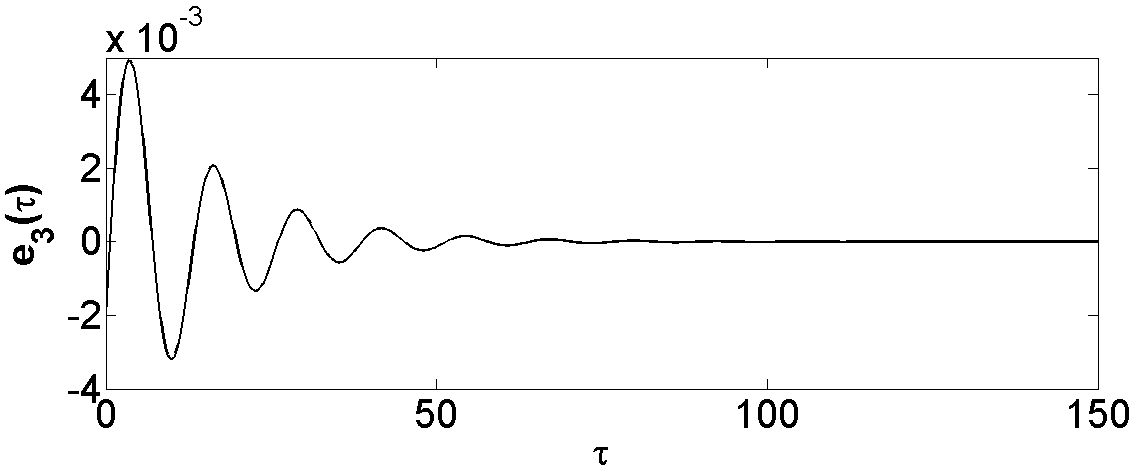}(e)
    \includegraphics[scale=0.22]{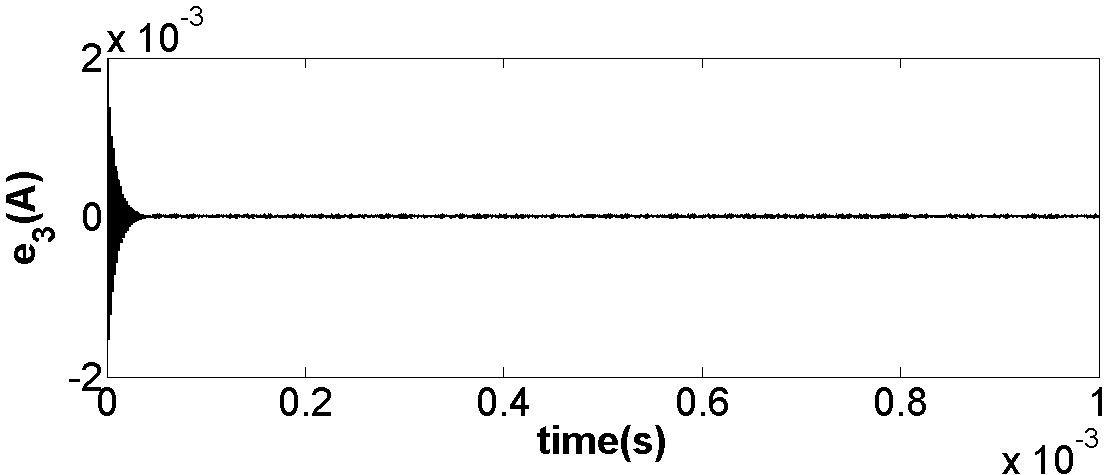}(f)
    \caption{Time evolution of the synchronization errors from Matlab simulations (Left) and Pspice simulations (Right).}
     \label{Fig.4}
     \end{center}
     \end{minipage}
     \end{figure}

The controller's parameters are $\varphi=10$ and $k=3$. The controller circuit was realized through the following relations: $k=\displaystyle\frac{R_5}{V_T~R_3}=\displaystyle\frac{R_6}{V_T~R_4}$ and $\varphi=R_7~I_s$ where $V_T\simeq 0.026Volt$ and $I_s\simeq 10^{-12}$ are diode characteristics. The Voltage controlled current source (VCCS) is used to minimize as much as possible the mutual influence of between the slave system (Green box) and the controller (Red box) and to only generate the current which obliges the response system to follow the drive system. The graphs of the of
Figs.~(\ref{Fig.3}) and (\ref{Fig.4}) show that the synchronization is reached around the
dimensionless time $\tau=60$.\\

\begin{figure}[htp]
 \begin{minipage}[b]{16cm}
      \begin{center}
    \includegraphics[scale=0.24]{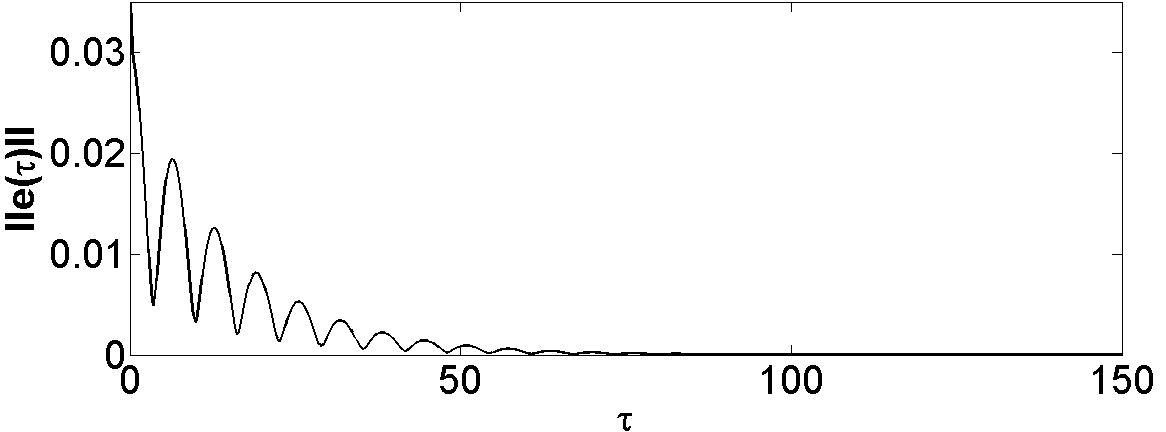}(a)
        \includegraphics[scale=0.24]{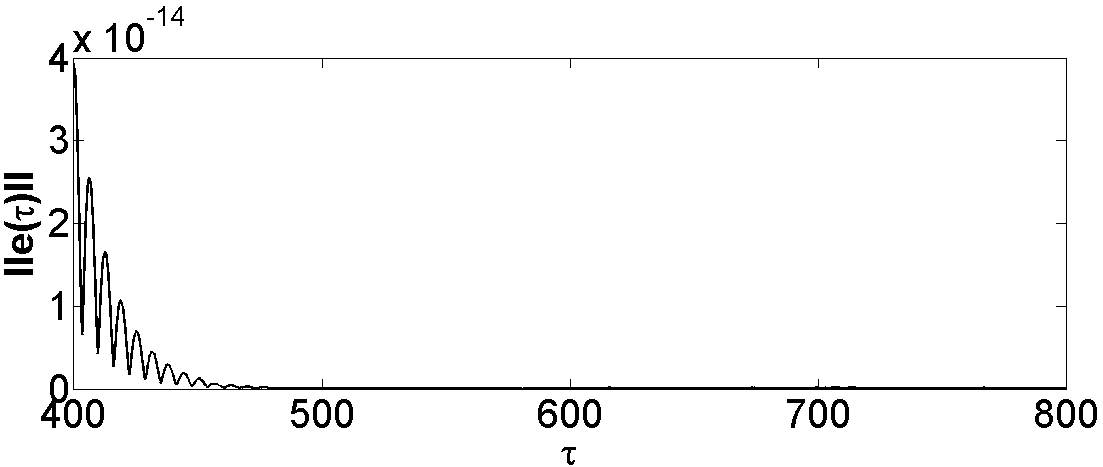}(b)
    \caption{Time evolution of the synchronization errors norm with the propose
scheme (6).}
     \label{Fig.5}
     \end{center}
     \end{minipage}
     \end{figure}

\begin{figure}[htp]
 \begin{minipage}[b]{16cm}
      \begin{center}
    \includegraphics[scale=0.24]{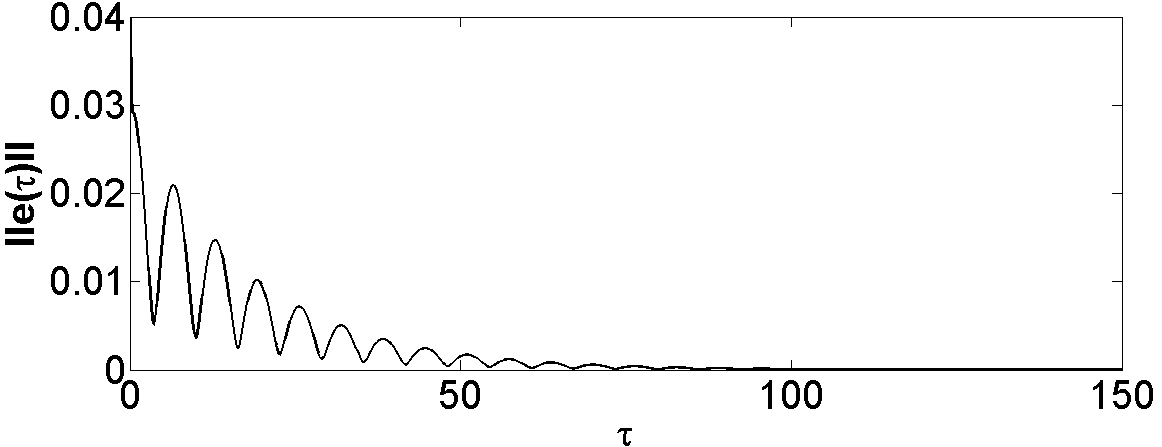}(a)
        \includegraphics[scale=0.24]{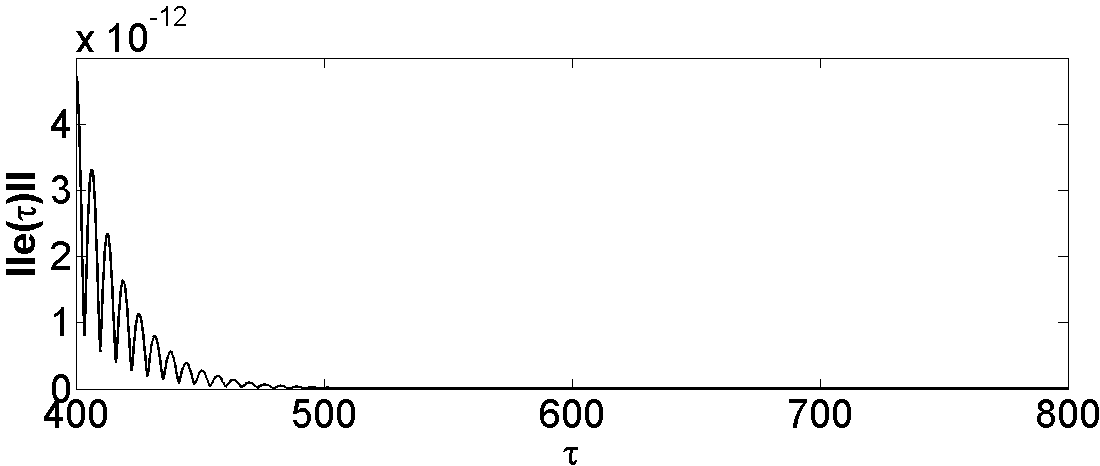}(b)
    \caption{Time evolution of the synchronization errors norm with linear
controller (16).}
     \label{Fig.6}
     \end{center}
     \end{minipage}
     \end{figure}

\begin{rmq}
In Pspice simulations, the synchronization is reached for a high values of $R_7$ particularly if $R_7 > 100k\Omega$. $R_7$ role is to increase the
value of the VCCS output current by increasing the value of the voltage at its landmarks.
\end{rmq}
Considering the case without disturbances, if we compare the proposed scheme
with the one for which the controller is given by the following relation,

\begin{equation} \label{16}
\begin{array}{lcl}
U(t)=-\zeta  e_1\left( \tau\right),
\end{array}
\end{equation}
where $\zeta$ is a positive constant chosen equal to $\varphi$, it appears
that, as one can visually appreciate on
the graphs of Figs~(\ref{Fig.5}) and (\ref{Fig.6}), the exponential function based nonlinear
controller is faster than the linear
controller with fixed gain.

\newpage
\section{Conclusion}
\noindent

In this paper the synchronization between two different delayed chaotic
systems is studied via a simple -- exponential function based -- nonlinear controller.
Although different initial conditions and disturbances make synchronization more difficulty, a simple exponential function based nonlinear controller is
designed which facilitates the task. This is proven through the Lyapunov stability theory, it is shown that both
master-slave systems should be practically synchronized. It is important
to note that the proposed scheme improves the linear controller with fixed gain
usually used. To show the effectiveness of the proposed strategy, some numerical
simulations are given, they show the efficiency of the proposed
strategy in front of the linear fixed gain based controller. The electronic
circuit of the used controller is also given followed by some simulations.

\section*{Acknowledgments}
\noindent

The authors thank the hospitality of the University of Yaounde I and the Abdus Salam International Centre for Theoretical Physics. HAC acknowledges support by the
CNPq-ProAfrica, project nr. 490265/2010-3 (Brazil). PL acknowledges M. Andr\'{e} Romet for support throughout this work.
\\

\section*{\textbf{References}}

  $[1]$  Pecora L M,  Carroll T L. Synchronization in chaotic systems. Phys Rev Lett 1990;64:821-4.\\
  $[2]$ Li X, Guan X and Ru D. The damping time of EEG with information retrieve and autoregressive
models. In the 5th IFAC symposium on Modelling and Control in
Biomedical Systems August (2003), Australia.\\
  $[3]$ Han S K C,   Kerrer C and   Kuramoto T. Dephasing and Bursting in coupled neural
  oscillators. Phys Rev Lett 1995;75:3190-3.\\
  $[4]$ Lin J, Huang C, Liao T and Yan J. Design and implementation of digital secure communication based on synchronized
  chaotic systems. Digital Signal Processing 2010;20:229-237.\\
$[5]$ Islam N, Islam B and  Mazumdar H P. Generalized chaos synchronization of unidirectionally
coupled Shimizu-Morioka dynamical systems. Differential Geometry - Dynamical Systems 2011;13:101-6.\\
$[6]$ Blasius B, Huppert A and Stone L. Complex dynamics and phase synchronization in spatially extended ecological systems.
Nature 1999;399:354-9.\\
$[7]$ Sivaprakasam S, Pierce I, Rees P, Spencer P S, Shore K A and Valle A. Inverse synchronization in
semiconductor laser diodes. Phys Rev A ;64:013805-1-8.\\
$[8]$ Wedekind I and Parlitz U. Synchronization and antisynchronization of chaotic power drop-outs
and jump-ups of coupled semiconductor lasers. Phys Rev E 2002;66:026218-1-4.\\
$[9]$ Fradkov A, Nijmeijer H and Markov A. Adaptive observer-based synchronization for
communication. International Journal of Bifurcation and Chaos 2000;10:2807-13.\\
 $[10]$ Cuomo K, Oppenheim A V. Circuit implementation of synchronized chaos with applications to communications. Phys Rev Lett
1993;71:658.\\
 $[11]$ Bowong S. Stability analysis for the synchronization of chaotic systems with different order:
 Application to secure communication. Phys Lett A 2004;326:102-13.\\
$[12]$ Bowong S, Tewa J J. Unknown inputs'adaptive observer for a class of chaotic systems with
uncertainties. Mathematical and Computer Modeling 2008;48:1826-39.\\
$[13]$ Fotsin H B and Bowong S. Adaptive control and synchronization of chaotic systems consisting
of Van der Pol oscillators coupled to linear oscillators. Chaos, Solitons \& Fractals 2006;27:822-35.\\
$[14]$ Astolfi A, Karagiannis D and Ortega R. Nonlinear and Adaptive Control with
Applications.  Springer-Verlag 2008.\\
 $[15]$ Feng G and Lozano R. Adaptive control systems. Reed Elsevier plc group First published 1999.\\
$[16]$ Shahverdiev E M, Nuriev R A, Hashimova L H, Huseynova E M a,
Hashimov R H and Shore K A. Complete inverse chaos synchronization,
parameter mismatches and generalized synchronization
in the multi-feedback Ikeda model. Chaos Solitons \& Fractals 2008;36:2116.\\
$[17]$ Sundarapandian V. Global chaos anti-Synchronization of Liu and Chen Systems by Nonlinear
Control. International Journal of Mathematical Sciences and Applications 2011;1:691-702.\\
 $[18]$  Zhang X and Zhu H. Anti-synchronisation of two different hyperchaotic systems via active and
adaptive control. International Journal of Nonlinear Science 2008;6:216-23.\\
$[19]$  Zhu H. Anti-synchronization of two different chaotic systems via optimal control
with fully unknown parameters. Journal of Information and Computing Science 2010;5:011-8.\\
$[20]$ Gao X, Zhony S and Gao F. Exponential synchronization of neural networks with time-varying
delays.   Nonlinear Analysis 2009;71:2003-11.\\
$[21]$  Zheng S, Bi Q and Cai G. Adaptive projective synchronization in complex networks with
time-varying  coupling delay. Phys Lett A 2009;373:1553-9.\\
$[22]$  Cai J, Lin M and Yuan Z. Secure communication using practical synchronization between
two different chaotic systems with uncertainties. Mathematical and Computational Application 2010;15: 166-75.\\
$[23]$  Louodop F P H, Fotsin H B and Bowong S. A strategy for adaptive synchronization
of an electrical chaotic circuit based
on nonlinear control. Physica Scripta 2012;85(025002):6pp.\\
$[24]$ Roopaei M and Argha A, 2011.  Novel Adaptive Sliding Mode Synchronization in a Class of Chaotic
Systems. World Applied Sciences Journal 2011;12:2210-17.\\
$[25]$  Sun Z and Yang X. Parameters identification and synchronization of chaotic delayed systems
containing uncertainties and time-varying delay. Mathematical Problems in Engineering 2010;2010:1-15.\\
$[26]$ Kammongne S T and Fotsin H B. Synchronization of modified Colpitts oscillators with
structural perturbations. Physical Scripta 2011;83 (065011):7pp.\\
$[27]$ Choon Ki Ahn. Robust chaos synchronization using input-to-state stable control.
Pramana Journal of Physics 2010;74:705-18.\\
$[28]$ Earn D J D, Rohani P and Grenfell B T. Persistence, chaos and synchrony in ecology
and epidemiology. Proc Royal Societ London B 1998;265:7-10.\\
$[29]$ Bowong S. Optimal control of the transmission dynamics of tuberculosis. Nonlinear Dynamics 2010;
DOI 10.1007/s11071-010-9683-9.\\
$[30]$ Yang Z, Pedersen G K M and Pedersen J H. Model-Based Control of a Nonlinear One
Dimensional Magnetic Levitation with a Permanent-Magnet Object. Automation and Robotics;21:359-74\\
$[31]$  Yuan-Zhao Yin. Experimental demonstration of chaotic synchronization in the modified Chua'oscillators.
International Journal of Bifurcation and Chaos 1997;7:1401-10.\\
$[32]$ Xiao-Xin L, Hai-Geng L, Gang Z, Ji-gui J, Xiao-Jun Z and Bing-Ji X. New results on
global synchronization of Chua's circuit. Acta Automatica Sinica 2005;31:320-6.\\
$[33]$ Chua L O. The genesis of Chua's circuit. Archiv Electronik Übertragungstechnik 1992;46:250-7.\\
$[34]$ Chua L O, Yang T, Zhong G and Wu C W. Synchronization of Chua's circuits with
time-varying channels  and parameters. IEEE Transactions on Circuits and Systems: Fundamental
Theory and Applications 1996;43:862-8.\\
$[35]$ Markov A Y, Fradkov A L and Simin G S. Adaptive synchronization of chaotic generators based on tunnel diodes.
 Proceeding of 35th Conference on Decision and Control December 1996;TM04:2177-82.\\
$[36]$ Ding Z and Cheng G. A new uniformly ultimate boundedness criterion for discrete-time nonlinear systems.
 Applied Mathematics 2011;2:1323-6.\\
$[37]$ De La Sen M and  Alonso S. Adaptive control of time-invariant systems
with discrete dynamics delays subject to multiestimation.
 Discrete Dynamics in Nature and Society 2006;Article ID 41973:1-27.\\
$[38]$ Bitsoris G, Vassilaki M, and Athanasopoulos N. Robust positive invariance and ultimate boundedness of nonlinear systems.
 Preprints of the 20th Mediterranean Conference on
Control and Automation (MED), Barcelona, Spain, July 3-6 2012;ThA1.2:598-603.

\end{document}